\documentclass[notitlepage,12pt]{revtex4-1}
\usepackage{amsmath}
\usepackage{slashed}
\usepackage{txfonts}
\usepackage{microtype}
\usepackage{graphicx}
\usepackage[
breaklinks=true
]{hyperref}
\usepackage{color}
\usepackage{isomath}
\def\sout{\bgroup\markoverwith
{\textcolor{red}{\rule[0.5ex]{2pt}{0.5pt}}}\ULon}
\def\be{\begin{equation}}
\def\ee{\end{equation}}
\def\bes{\begin{equation*}}
\def\ees{\end{equation*}}
\def\bea{\begin{eqnarray}}
\def\eea{\end{eqnarray}}
\def\beas{\begin{eqnarray*}}
\def\eeas{\end{eqnarray*}}
\def\bal#1\eal{\begin{align}#1\end{align}}
\def\bals#1\eals{\begin{align*}#1\end{align*}}

\newcommand{\ket}[1]{|#1\rangle}
\newcommand{\braket}[2]{\langle #1|#2\rangle}

\renewcommand{\vec}[1]{\mathbf{#1}} 
\newcommand{\del}{\partial}

\graphicspath{{figures/}}

\renewcommand*{\vec}[1]{\boldsymbol{#1}}

\begin{document}
\title{Volkov wave function: its orthonormality and completeness}
\author{Enderalp Yakaboylu}

\begin{abstract}
The present note aims to provide a clear and explicit derivation of the orthonormality condition, and the completeness property of the Volkov wave function. Thus, none of the results are new.
\end{abstract}

\maketitle

\section{Volkov wave function}

The Dirac equation for a charged particle interacting with an arbitrary electromagnetic field is written as
\be
\label{Dirac_ket_eqn}
i \frac{d \ket{\psi(t)}}{d t} = \left[ \vec{\alpha} \cdot (\vec{p} - e \vec{A}) + e \phi + m \beta \right] \ket{\psi(t)} \, ,
\ee
where we used the natural units such that $\hbar = c = 1$, $e$ is the charge of the particle, $\vec{\alpha}$ and $\beta$ are the Dirac matrices chosen in the standard representation~\cite{Bjorken1964relativistic}, and the gauge potential is given by $A^\mu = (\phi, \vec{A})$ with the metric convention $(+,-,-,-)$. For the position representation of the state vector the Dirac equation~(\ref{Dirac_ket_eqn}) reads 
\be
\label{Dirac_func_eqn}
\left[ i \gamma^\mu \left(\del_\mu + i e A_\mu \right) - m \right] \braket{\vec{x}}{\psi (t)} = 0 \, ,
\ee
with $\gamma^\mu = (\beta, \beta \, \vec{\alpha})$. 

If we multiple Eq.~(\ref{Dirac_func_eqn}) with the operator $i \gamma^\nu D_\nu + m$ from the left side, we provide
\be
\label{Dirac_1}
 0 = \left[ i \gamma^\nu D_\nu + m \right] \left[ i \gamma^\mu D_\mu - m \right] \psi (\vec{x},t) = \left[ D^2 + \frac{\gamma^\nu \gamma^\mu}{2} [D_\nu, D_\mu] + m^2 \right] \psi (\vec{x},t) \, ,
\ee
with $D_\mu = \del_\mu + i e A_\mu$ being the covariant derivative, $\psi (\vec{x},t) \equiv \braket{\vec{x}}{\psi (t)} $, and we used the commutation relation $\left\{ \gamma^\nu, \gamma^\mu \right\} = 2 g^{\nu \mu}$. Using the fact that $[D_\nu, D_\mu] = i e F_{\nu \mu}$, where $F_{\nu \mu}  = \del_\nu A_\mu - \del_\mu A_\nu$ is the field strength tensor, Eq.~(\ref{Dirac_1}) can be further written as
\be
\label{Dirac_2}
\left[\del^2  + 2 i e A^\mu \del_\mu -e^2 A^2 + \frac{i e}{2} \gamma^\nu \gamma^\mu F_{\nu \mu} + m^2 \right] \psi (\vec{x},t) = 0 \, ,
\ee
where we have used the Lorenz gauge condition $\del_\mu A^\mu = 0$. We observe from Eq.~(\ref{Dirac_2}) that as the gamma matrices couple to the field strength tensor, the spinor part of the wave function is gauge invariant for an arbitrary electromagnetic field.

Our point of interest is the solution of Eq.~(\ref{Dirac_2}) for a plane wave. A plane wave can be defined in terms of the following gauge potential $A^\mu = A^\mu (\eta)$ with the phase $\eta = k^\mu x_\mu$, where $k^\mu$ is the wave vector fulfilling the condition $k^2 = 0$. For a plane wave the spin resolved term of Eq.~(\ref{Dirac_2}) can be given by
\be
\frac{i e}{2} \gamma^\nu \gamma^\mu F_{\nu \mu} =  i e \gamma^\nu \gamma^\mu \del_\nu A_\mu =  i e \gamma^\nu \gamma^\mu \dot{A}_\mu k_\nu \, ,
\ee
where the dot denotes derivative with respect to the gauge field's argument $\eta$. Now let us seek the solution of Eq.~(\ref{Dirac_2}) in the form of 
\be
\label{anstaz}
\psi^{(\pm)} (\vec{x},t) =  \exp\left( -i  (\pm p^\mu) x_\mu  \right) f^{(\pm)}(\eta) u_{p,s}^{(\pm)} \, ,
\ee
where $(+)$ and $(-)$ label the particle and anti-particle solutions, respectively. The free particle/anti-particle spinor with the normalization ${u_{p,s'}^{(\pm)}}^{\dagger} u_{p,s}^{(\pm)} = \delta_{s,s'}$  can be given by
\be
u_{p,s}^{(+)} = \sqrt{\frac{p^0 + m}{2 p^0}} \begin{pmatrix}
            \phi_s \\
            \displaystyle \frac{\vec{p} \cdot \vec{\sigma} \, \phi_s}{p^0 + m}  \\
          \end{pmatrix} \, , \quad u_{p,s}^{(-)} = \sqrt{\frac{p^0 + m}{2 p^0}} \begin{pmatrix}
            \displaystyle \frac{\vec{p} \cdot \vec{\sigma} \, \chi_s}{p^0 + m} \\
             \chi_s  \\
          \end{pmatrix} \, ,
\ee
where $p^\mu = (p^0,\vec{p})$ is the free particle/anti-particle's four-momentum with the energy $p^0 = \sqrt{m^2 + \vec{p}^2} > 0$, $\phi_1 = \chi_2 = (1 \quad 0)^T$, $\phi_2 = \chi_1 = (0 \quad 1)^T$ are two component spinors, and $\vec{\sigma}$ are the Pauli matrices \cite{Bjorken1964relativistic}. In the absence of the plane wave, $f^{(\pm)}(\eta)$ reduces to the identity, and $\psi^{(\pm)} (\vec{x},t) $ becomes the field-free solution of the Dirac equation. If we plug the ansatz~(\ref{anstaz}) into Eq.~(\ref{Dirac_2}), the unknown function $f^{(\pm)}(\eta)$ satisfies the following equation
\bal
\label{Dirac_3}
\nonumber & \left[-p^2 f^{(\pm)}(\eta) \mp 2 i p^\mu k_\mu \dot{f}^{(\pm)}(\eta) + 2ie A^\mu \left(\mp  i p_\mu f^{(\pm)}(\eta) +  k_\mu \dot{f}^{(\pm)}(\eta)\right) \right. \\
& \left. - e^2 A^2 f^{(\pm)}(\eta)+i e \gamma^\mu k_\mu \gamma^\nu \dot{A}_\nu f^{(\pm)}(\eta) + m^2 f^{(\pm)}(\eta) \right] u_{p,s}^{(\pm)} = 0 \, .
\eal
Before going further, we first observe that the Lorenz gauge condition for a plane wave can also be written as $\del_\mu A^\mu = \dot{A}^\mu k_\mu = d (A^\mu k_\mu)/ d \eta $, which implies $A^\mu k_\mu$ equals to a constant, and in fact, without loss of generality, this constant can be set to zero. Using further the on-shell condition $p^2 = m^2$, Eq.~\ref{Dirac_3} simplifies to
\be
\label{Dirac_4}
\left[ \mp 2 i p^\mu k_\mu \dot{f}^{(\pm)}(\eta) +  \left( \pm 2e A^\mu p_\mu - e^2 A^2 + i e \gamma^\mu k_\mu \gamma^\nu \dot{A}_\nu \right) f^{(\pm)}(\eta)  \right] u_{p,s}^{(\pm)} = 0 \, ,
\ee
and whose solution just yields
\be
f^{(\pm)}(\eta) = \exp\left(\mp i \frac{e}{ pk} \int^\eta_{\eta_0} d \eta' \left( \pm A p - e\frac{A^2}{2}\right) \right) \exp\left( \pm e \frac{\slashed{k}\slashed{A}}{2 pk} \right) \, ,
\ee
where $\slashed{k} \equiv \gamma^\mu k_\mu$, and $\eta_0$ is the initial phase at which the gauge potential and the field vanish. We further notice that
\be
\exp\left(\pm e \frac{\slashed{k}\slashed{A}}{2 pk} \right) = 1 \pm e \frac{\slashed{k}\slashed{A}}{2 pk} \, ,
\ee
due to the fact that $k^2 =0$. As a result, the wave function with the quantum numbers $p$ and $s$ reads
\be
\label{Volkov}
\psi_{p,s}^{(\pm)} (\vec{x},t) =  \exp\left[ -i \left( \pm p^\mu x_\mu  \pm  \frac{e}{ pk} \int^\eta_{\eta_0} d\eta' \, \left( \pm A p - e A^2/2\right) \right)\right] \left( 1 \pm e \frac{\slashed{k}\slashed{A}}{2 pk}  \right) u_{p,s}^{(\pm)} \, ,
\ee
which is called the Volkov wave function~\cite{Volkov_1935}. 

\section{orthonormality of Volkov wave function}

In this section, following Ritus~\cite{Ritus1985quantum}, we explicitly show that the Volkov wave function obeys the usual orthonormality condition
\be
\label{orthonormality_1}
\int d^3 \vec{x} \,  {\psi_{p', s'}^{(\pm)}}^{\dagger} (\vec{x},t) \psi_{p, s}^{(\pm)} (\vec{x},t) =  \delta(\vec{p}'-\vec{p})\delta_{s',s} \, .
\ee

For a shorthand notation, we first rewrite the Volkov wave function in the following form
\be
\label{Volkov_1}
\psi_{p,s}^{(\pm)} (\vec{x},t) =  \exp\left[ - i \left( \pm S_p^{(\pm)} \right) \right] \left( 1 \pm e \frac{\slashed{k}\slashed{A}}{2 pk}  \right) u_{p,s}^{(\pm)} \, ,
\ee
with the phase
\be
S_p^{(\pm)} \equiv p^\mu x_\mu  +  \frac{e}{ pk} \int^\eta_{\eta_0} d\eta' \, \left( \pm A p - e A^2/2\right) \, .
\ee
Then, we set the scalar part of the gauge potential zero within the Lorenz gauge, and choose the propagation direction along the $z$-direction and the vector potential on the $x-y$ plane, i.e.,
\be
A^\mu (\eta) = \left(0, A_x (t-z), A_y (t-z), 0 \right) \,.
\ee
Thereby, the phase of the Volkov wave function is
\be
S_p^{(\pm)} = p^0 t - p_z z - \vec{p}_\perp \cdot \vec{x} +  \int_0^{t-z} d \eta' \, \frac{\mp 2 e \vec{p}_\perp \cdot \vec{A} + e^2 \vec{A^2} }{2 p_{-}}  \, ,
\ee
where for the sake of convenience we set $\eta_0  = 0 $, $\vec{p}_\perp = (p_x,p_y)$, and $p_{-} = p^\mu k_\mu = p^0 - p_z >0$. The phase can also be rewritten as
\be
\label{phase_2}
S_p^{(\pm)} = p^0 t - p_z z - \vec{p}_\perp \cdot \vec{x} +  \int_0^{t-z} d \eta' \, \frac{\left( \vec{p}_\perp \mp e \vec{A} \right)^2 -\vec{p}_\perp^2 }{2 p_{-}} = p_{-} \frac{t + z}{2} - \vec{p}_\perp \cdot \vec{x} +  \int_0^{t-z} d \eta' \, \frac{\left( \vec{p}_\perp \mp e \vec{A} \right)^2 + m^2 }{2 p_{-}} \, .
\ee

As the integrand is independent from $x$ and $y$, one can immediately evaluate the corresponding integrals in Eq.~\ref{orthonormality_1}, which are in fact Dirac delta functions. Therefore,  the orthonormality reads
\bal
& \int d^3 \vec{x} \,  {\psi_{p', s'}^{(\pm)}}^{\dagger} (\vec{x},t) \psi_{p, s}^{(\pm)} (\vec{x},t) = \int d^3 \vec{x} \, \exp\left[ \pm i  \left( S_{p'}^{(\pm)} -  S_p^{(\pm)}  \right) \right] {U_{p', s'}^{(\pm)}}^{\dagger} U_{p, s}^{(\pm)} \\
\nonumber & = \delta(\vec{p'}_\perp - \vec{p}_\perp) \int dz \, \exp\left[ \pm i \left( (p'_{-} - p_{-} ) \frac{t + z}{2} -  (p'_{-}-p_{-}) \int_0^{t-z} d \eta' \, \frac{\left( \vec{p}_\perp \mp e \vec{A} \right)^2 + m^2 }{2 p_{-} p'_{-}} \right) \right]   {U_{p', s'}^{(\pm)}}^{\dagger} U_{p, s}^{(\pm)} \, ,
\eal
with
\be
U_{p, s}^{(\pm)} \equiv \left( 1 \pm e \frac{\slashed{k}\slashed{A}}{2 p_{-}}  \right) u_{p,s}^{(\pm)} \, .
\ee
Now, we define
\be
\zeta^{(\pm)} = z -  \int_0^{t-z} d \eta' \, \frac{\left( \vec{p}_\perp \mp e \vec{A} \right)^2 + m^2 }{p_{-} p'_{-}} \, ,
\ee
which implies
\be
d \zeta^{(\pm)}  = d z \left( 1 + \frac{\left( \vec{p}_\perp \mp e \vec{A} \right)^2 + m^2 }{p_{-} p'_{-}} \right) \,, 
\ee
and hence simplifies the integral to
\bal
\label{normalization}
\nonumber & \int d^3 \vec{x} \,  {\psi_{p', s'}^{(\pm)}}^{\dagger} (\vec{x},t) \psi_{p, s}^{(\pm)} (\vec{x},t) = \delta(\vec{p'}_\perp - \vec{p}_\perp) \\
& \times \int_{-\infty}^{\infty} d \zeta^{(\pm)}  \frac{{U_{p', s'}^{(\pm)}}^{\dagger} U_{p, s}^{(\pm)} }{1 + \left[ \left( \vec{p}_\perp \mp  e \vec{A} \right)^2 + m^2 \right]/ p_{-} p'_{-} } \exp\left[ \pm i (p'_{-} - p_{-} ) \frac{t + \zeta}{2} \right] \, .
\eal
The spin resolved part can be written as
\be
{U_{p', s'}^{(\pm)}}^{\dagger} U_{p, s}^{(\pm)} ={u_{p',s'}^{(\pm)}}^\dagger \left( 1 \pm e \frac{\gamma^0 \slashed{A} \slashed{k} \gamma^0}{2 p'_{-}}  \right) \left( 1 \pm e \frac{\slashed{k}\slashed{A}}{2 p_{-}}  \right) u_{p,s}^{(\pm)} \, ,
\ee
and the intermediate square matrix can be decomposed as
\bal
 \left( 1 \pm e \frac{\gamma^0 \slashed{A} \slashed{k} \gamma^0}{2 p'_{-}}  \right) \left( 1 \pm e \frac{\slashed{k}\slashed{A}}{2 p_{-}}  \right) & = 1 + \frac{1}{p'_{-} p_{-}} \left( e^2 \vec{A}^2 \frac{\gamma^0 (\gamma^0-\gamma^3)}{2} \mp 2 e A_x \gamma^1 \frac{(p'_{-}-p_{-}) \gamma^3 - (p'_{-}+p_{-}) \gamma^0}{4} \right. \\
& \left. \mp 2 e A_y \gamma^2 \frac{(p'_{-}-p_{-}) \gamma^3 - (p'_{-}+p_{-}) \gamma^0}{4}  \right) \, .
\eal
If we evaluate each corresponding term, we provide the following expressions
\bal
{u_{p',s'}^{(\pm)}}^\dagger u_{p,s}^{(\pm)} & =  \left( 1 + \frac{\vec{p}_\perp^2 + m^2}{p'_{-} p_{-}} \right) g(p',p)  \delta_{s',s} \, ,\\
{u_{p',s'}^{(\pm)}}^\dagger \frac{\gamma^0 (\gamma^0-\gamma^3)}{2} u_{p,s}^{(\pm)}  & =g(p',p)  \delta_{s',s} \, , \\
{u_{p',s'}^{(\pm)}}^\dagger \gamma^1 \frac{(p'_{-}-p_{-}) \gamma^3 - (p'_{-}+p_{-}) \gamma^0}{4} u_{p,s}^{(\pm)}  & =   p_x \,  g(p',p)  \delta_{s',s} \, , \\
{u_{p',s'}^{(\pm)}}^\dagger \gamma^2 \frac{(p'_{-}-p_{-}) \gamma^3 - (p'_{-}+p_{-}) \gamma^0}{4} u_{p,s}^{(\pm)}  & =   p_y \, g(p',p)  \delta_{s',s} \, ,
\eal
where the common function
\be
g(p',p) = \frac{\left(p'_{-} + m\right)\left(p_{-} + m\right) + \vec{p}_\perp^2}{4 \sqrt{(m+p^0)(m+p'^0)p^0 p'^0}} \ .
\ee
As a result, we obtain
\be
{U_{p', s'}^{(\pm)}}^{\dagger} U_{p, s}^{(\pm)} =  \left( 1 + \left[ \left( \vec{p}_\perp \mp e \vec{A} \right)^2 + m^2 \right]/ p_{-} p'_{-} \right)g(p',p) \delta_{s',s} \, ,
\ee
which exactly cancels the corresponding denominator in Eq.~(\ref{normalization}), and therefore we find
\be
\int d^3 \vec{x} \,  {\psi_{p', s'}^{(\pm)}}^{\dagger} (\vec{x},t) \psi_{p, s}^{(\pm)} (\vec{x},t) = \delta(\vec{p'}_\perp - \vec{p}_\perp) \delta\left(\frac{p'_{-} - p_{-}}{2}\right) g(p,p') \delta_{s',s} \, .
\ee
The function $g(p,p')$ reduces to
\be
g(p,p) = \frac{p_{-}}{2 p^0}
\ee
for the same momentum. Finally, using the relation
\be
 \frac{p_{-}}{2 p^0} \delta\left(\frac{p'_{-} - p_{-}}{2}\right) = \delta(p'_z - p_z) \, ,
\ee
we show that 
\be
\int d^3 \vec{x} \,  {\psi_{p', s'}^{(\pm)}}^{\dagger} (\vec{x},t) \psi_{p, s}^{(\pm)} (\vec{x},t) =  \delta(\vec{p}'-\vec{p})\delta_{s',s} \, .
\ee

\section{completeness of Volkov wave function}

In this section, we present a clear derivation of the completeness property of the Volkov wave function, explicitly we will show that
\be
\mathcal{C}(\vec{x},\vec{x}') = \sum_{s = 1}^2 \int d^3 \vec{p} \left(  \psi_{p, s}^{(+)} (\vec{x},t) {\psi_{p, s}^{(+)}}^{\dagger} (\vec{x}',t)  + \psi_{p, s}^{(-)} (\vec{x},t) {\psi_{p, s}^{(-)}}^{\dagger} (\vec{x}',t)\right) = \delta(\vec{x}-\vec{x}') \, . 
\ee

Using the form of the Volkov wave function given in Eq.~(\ref{Volkov_1}), the completeness can be given by
\bal
\label{completeness_1}
\nonumber \mathcal{C}(\vec{x},\vec{x}') & = \int d^3 \vec{p} \exp\left[ -i S^{(+)} (\vec{x}) + i S^{(+)} (\vec{x}') \right]  U_{\vec{p}, s}^{(+)}(\vec{x}) {U_{\vec{p}, s}^{(+)}}^{\dagger} (\vec{x}')  \\
& + \int d^3 \vec{p} \exp\left[ i S^{(-)} (\vec{x}) - i S^{(-)} (\vec{x}') \right] U_{\vec{p}, s}^{(-)}(\vec{x}) {U_{\vec{p}, s}^{(-)}}^{\dagger} (\vec{x}') \, ,
\eal
with
\bal
 U_{\vec{p}, s}^{(\pm)}(\vec{x}) {U_{\vec{p}, s}^{(\pm)}}^{\dagger} (\vec{x}') & =  \left( 1 \pm e \frac{\slashed{k}\slashed{A}(\vec{x})}{2 p_{-}}  \right)  \sum_s  u_{p,s}^{(\pm)} {u_{p,s}^{(\pm)}}^\dagger \left( 1 \pm e \frac{\gamma^0 \slashed{A}(\vec{x}') \slashed{k} \gamma^0}{2 p'_{-}}  \right) \\
 & = \left( 1 \pm e \frac{\slashed{k}\slashed{A}(\vec{x})}{2 p_{-}}  \right) \frac{\slashed{p} \pm m }{2 p^0} \gamma^0 \left( 1 \pm e \frac{\gamma^0 \slashed{A}(\vec{x}') \slashed{k} \gamma^0}{2 p'_{-}}  \right) \, ,
\eal
where we used the completeness of the free particle/anti-particle spinor in the second line. With the help of the phase~(\ref{phase_2}), Eq.~(\ref{completeness_1}) is rewritten as
\bal
\label{completeness_2}
\nonumber \mathcal{C}(\vec{x},\vec{x}') & = \int d^3 \vec{p} \,  e^{ \displaystyle -i \left[ \frac{p_{-}}{2} (z-z') - \vec{p}_\perp  \cdot (\vec{x} - \vec{x}') - \int_{t-z}^{t-z'} \frac{ \left( \vec{p}_\perp - e \vec{A} \right)^2 + m^2}{2 p_{-}} \right] } 
U_{\vec{p}, s}^{(+)}(\vec{x}) {U_{\vec{p}, s}^{(+)}}^{\dagger} (\vec{x}')  \\
& + \int d^3 \vec{p} \, e^{\displaystyle i\left[ \frac{p_{-}}{2} (z-z') - \vec{p}_\perp  \cdot (\vec{x} - \vec{x}') - \int_{t-z}^{t-z'} \frac{ \left( \vec{p}_\perp + e \vec{A} \right)^2 + m^2}{2 p_{-}} \right] } U_{\vec{p}, s}^{(-)}(\vec{x}) {U_{\vec{p}, s}^{(-)}}^{\dagger} (\vec{x}') \, .
\eal
After the following transformation $\vec{p} \rightarrow - \vec{p}$ in the second integral, we provide
\bal
\label{completeness_3}
\nonumber \mathcal{C}(\vec{x},\vec{x}') & = \int d^3 \vec{p} \,  e^{ \displaystyle -i \left[ \frac{p_{-}}{2} (z-z') - \vec{p}_\perp  \cdot (\vec{x} - \vec{x}') - \int_{t-z}^{t-z'} \frac{ \left( \vec{p}_\perp - e \vec{A} \right)^2 + m^2}{2 p_{-}} \right] } 
U_{\vec{p}, s}^{(+)}(\vec{x}) {U_{\vec{p}, s}^{(+)}}^{\dagger} (\vec{x}')  \\
& + \int d^3 \vec{p} \, e^{\displaystyle - i\left[ - \frac{p_{+}}{2} (z-z') - \vec{p}_\perp  \cdot (\vec{x} - \vec{x}') + \int_{t-z}^{t-z'} \frac{ \left( \vec{p}_\perp - e \vec{A} \right)^2 + m^2}{2 p_{+}} \right] } U_{-\vec{p}, s}^{(-)}(\vec{x}) {U_{-\vec{p}, s}^{(-)}}^{\dagger} (\vec{x}') \, ,
\eal
with $p_{+} = p^0 + p_z $. Now, for the first integral we apply the coordinate transformation $(\vec{p}) \rightarrow (\vec{p}_\perp, p_{-})  $, and for the second integral  $(\vec{p}) \rightarrow (\vec{p}_\perp, p_{+})  $. In other words, for the first integral
\be
p_z = \frac{m^2 + \vec{p}_\perp^2 - p_{-}^2 }{2 p_{-} } \,  \rightarrow \, \int_{-\infty}^{\infty} d p_{z} = \int_{0}^{\infty}  d p_{-} \left( \frac{m^2 + \vec{p}_\perp^2 + p_{-}^2 }{2 p_{-}^2 } \right) \, ,
\ee
and for the second integral
\be
p_z = \frac{m^2 + \vec{p}_\perp^2 - p_{+}^2 }{- 2 p_{+} } \,  \rightarrow \, \int_{-\infty}^{\infty}  d p_{z} = \int_{0}^{\infty}  d p_{+} \left( \frac{m^2 + \vec{p}_\perp^2 + p_{+}^2 }{2 p_{+}^2 } \right) \, .
\ee
Then, the completeness property becomes
\bal
\label{completeness_4}
\nonumber & \mathcal{C}(\vec{x},\vec{x}') = \\
\nonumber & \int d^2 \vec{p}_\perp \int_{0}^{\infty}  d p_{-} \left( \frac{m^2 + \vec{p}_\perp^2 + p_{-}^2 }{2 p_{-}^2 } \right) \,  e^{ \displaystyle -i \left[ \frac{p_{-}}{2} (z-z') - \vec{p}_\perp  \cdot  (\vec{x} - \vec{x}') - \int_{t-z}^{t-z'} \frac{ \left( \vec{p}_\perp - e \vec{A} \right)^2 + m^2}{2 p_{-}} \right] } \\
\nonumber & \times U_{\vec{p}, s}^{(+)}(\vec{x}) {U_{\vec{p}, s}^{(+)}}^{\dagger} (\vec{x}')  \\
\nonumber & + \int d^2 \vec{p}_\perp \int_{0}^{\infty}  d p_{+} \left( \frac{m^2 + \vec{p}_\perp^2 + p_{+}^2 }{2 p_{+}^2 } \right) \, e^{\displaystyle - i\left[ - \frac{p_{+}}{2} (z-z') - \vec{p}_\perp  \cdot  (\vec{x} - \vec{x}') + \int_{t-z}^{t-z'} \frac{ \left( \vec{p}_\perp - e \vec{A} \right)^2 + m^2}{2 p_{+}} \right] } \\
& \times U_{-\vec{p}, s}^{(-)}(\vec{x}) {U_{-\vec{p}, s}^{(-)}}^{\dagger} (\vec{x}') \, .
\eal
If we further apply the transformation $p_{+} \rightarrow - p_{-}$ in the second integral, we provide
\bal
\label{completeness_5}
\nonumber & \mathcal{C}(\vec{x},\vec{x}') = \\
\nonumber & \int d^2 \vec{p}_\perp \int_{0}^{\infty}  d p_{-} \left( \frac{m^2 + \vec{p}_\perp^2 + p_{-}^2 }{2 p_{-}^2 } \right) \,  e^{ \displaystyle -i \left[ \frac{p_{-}}{2} (z-z') - \vec{p}_\perp  \cdot (\vec{x} - \vec{x}') - \int_{t-z}^{t-z'} \frac{ \left( \vec{p}_\perp - e \vec{A} \right)^2 + m^2}{2 p_{-}} \right] } \\
\nonumber & \times U_{\vec{p}, s}^{(+)}(\vec{x}) {U_{\vec{p}, s}^{(+)}}^{\dagger} (\vec{x}')  \\
\nonumber & + \int d^2 \vec{p}_\perp \int_{-\infty}^{0}  d p_{-} \left( \frac{m^2 + \vec{p}_\perp^2 + p_{-}^2 }{2 p_{-}^2 } \right) \, e^{\displaystyle - i\left[ \frac{p_{-}}{2} (z-z') - \vec{p}_\perp  \cdot (\vec{x} - \vec{x}') - \int_{t-z}^{t-z'} \frac{ \left( \vec{p}_\perp - e \vec{A} \right)^2 + m^2}{2 p_{-}} \right] } \\
& \times \left. U_{-\vec{p}, s}^{(-)}(\vec{x}) {U_{-\vec{p}, s}^{(-)}}^{\dagger} (\vec{x}') \right|_{p_{+} = - p_{-}}\, .
\eal
At this level we observe that
\be
\label{spin_resolved_term}
U_{\vec{p}, s}^{(+)}(\vec{x}) {U_{\vec{p}, s}^{(+)}}^{\dagger} (\vec{x}') = \left. U_{-\vec{p}, s}^{(-)}(\vec{x}) {U_{-\vec{p}, s}^{(-)}}^{\dagger} (\vec{x}') \right|_{p_{+} = - p_{-}} \, ,
\ee
therefore, the completeness just yields
\be
\label{completeness_6}
\mathcal{C}(\vec{x},\vec{x}') = \int d^2 \vec{p}_\perp \exp\left( i \vec{p}_\perp  \cdot (\vec{x} - \vec{x}' \right) \int_{-\infty}^{\infty}  d p_{-} \,  \exp\left[  -i \left( \kappa_1 p_{-}  - \frac{\kappa_2}{p_{-}} \right) \right] \frac{p^0}{p_{-}}  U_{\vec{p}, s}^{(+)}(\vec{x}) {U_{\vec{p}, s}^{(+)}}^{\dagger} (\vec{x}') \, ,
\ee
where we used $ p^0 = \left( m^2 + \vec{p}_\perp^2 + p_{-}^2 \right)/ ( 2 p_{-}) $, and
\be
\kappa_1 \equiv \frac{z-z'}{2} \, , \quad \kappa_2 \equiv \int_{t-z}^{t-z'} \frac{ \left( \vec{p}_\perp - e \vec{A} \right)^2 + m^2}{2} \, 
\ee
such that $\kappa_1 \kappa_2 \ge 0$.

The spin resolved part~(\ref{spin_resolved_term}), on the other hand, can be written as
\bal
&\frac{p^0}{p_{-}}  U_{\vec{p}, s}^{(+)}(\vec{x}) {U_{\vec{p}, s}^{(+)}}^{\dagger} (\vec{x}') = \\
\nonumber & \frac{1}{2 p_{-}} \left\{ \slashed{p} + m + \frac{e}{2 p_{-}} \left[ ( \slashed{A}(z') \slashed{k} - \slashed{A}(z) \slashed{k} ) (\slashed{p} + m) - 2 p_{-} \slashed{A}(z') + 2 p A(z') \slashed{k} - e \slashed{k} \slashed{A}(z) \slashed{A}(z') \right] \right\} \gamma^0 \, .
\eal
We notice that the above equation is in the form of
\be
\frac{p^0}{p_{-}}  U_{\vec{p}, s}^{(+)}(\vec{x}) {U_{\vec{p}, s}^{(+)}}^{\dagger} (\vec{x}') = K_0 + \frac{K_1}{p_{-}} + \frac{K_2}{p_{-}^2} \, ,
\ee
where
\begin{subequations}
\bal
K_0 & = \frac{\gamma^0+\gamma^3}{4}\gamma^0  \, , \\
K_1 & = \frac{1}{2} \left[ m - \vec{p}_\perp  \cdot \vec{\gamma} + \frac{e}{2} \left[ \slashed{A}(z')- \slashed{A}(z) \right] \gamma^0 (\gamma^0 + \gamma^3)- e  \slashed{A}(z') \right]\gamma^0  \, , \\
K_2 & =  \frac{1}{4} \left[  m^2 + \vec{p}_\perp^2 + e\left[  \slashed{A}(z')- \slashed{A}(z)\right] \vec{p}_\perp  \cdot \vec{\gamma}  + e \, m \left[ \slashed{A}(z') - \slashed{A}(z)  \right]  - e 2 \vec{p}_\perp \cdot \vec{A}(z) -  e^2  \slashed{A}(z) \slashed{A}(z') \right] \slashed{k} \gamma^0 \,.
\eal 
\end{subequations}
Finally, using the expression (see Appendix~\ref{contour_integral})
\be
\int_{-\infty}^{\infty}  d p_{-} \,  \frac{\exp\left[  -i \left( \kappa_1 p_{-}  - \dfrac{\kappa_2}{p_{-}} \right) \right]}{p_{-}^n} = \begin{cases} 
      \delta(\kappa_1) & n = 0 \\
      0 & n = 1 \\
      \delta(\kappa_2) & n = 2
   \end{cases} \, ,
\ee
with
\be
\delta(\kappa_1) = 2 \delta(z-z')  \, , \quad \delta(\kappa_2) = \frac{2 \delta(z-z')}{\left( \vec{p}_\perp - e \vec{A} \right)^2 + m^2} \, , 
\ee
we prove that the completeness property can be written as
\bal
\label{completeness_7}
\nonumber \mathcal{C}(\vec{x},\vec{x}') & = \int d^2 \vec{p}_\perp \exp\left( i \vec{p}_\perp \cdot (\vec{x} - \vec{x}' \right) \left[ \frac{\delta(\kappa_2)}{4} \left( \left( \vec{p}_\perp - e \vec{A} \right)^2 + m^2 \right) (\gamma^0 - \gamma^3) + \frac{\delta(\kappa_1)}{4} (\gamma^0 +\gamma^3)\right] \gamma^0 \\
& =  \delta(\vec{x}-\vec{x}')\, .
\eal

\section*{Acknowledgments}

I would like to thank Karen Z. Hatsagortsyan and O. Skoromnik for giving me motivation to write this note. 

\section{Appendix} \label{contour_integral}

\begin{figure}[b]
  \centering
  \includegraphics[width=0.6\linewidth]{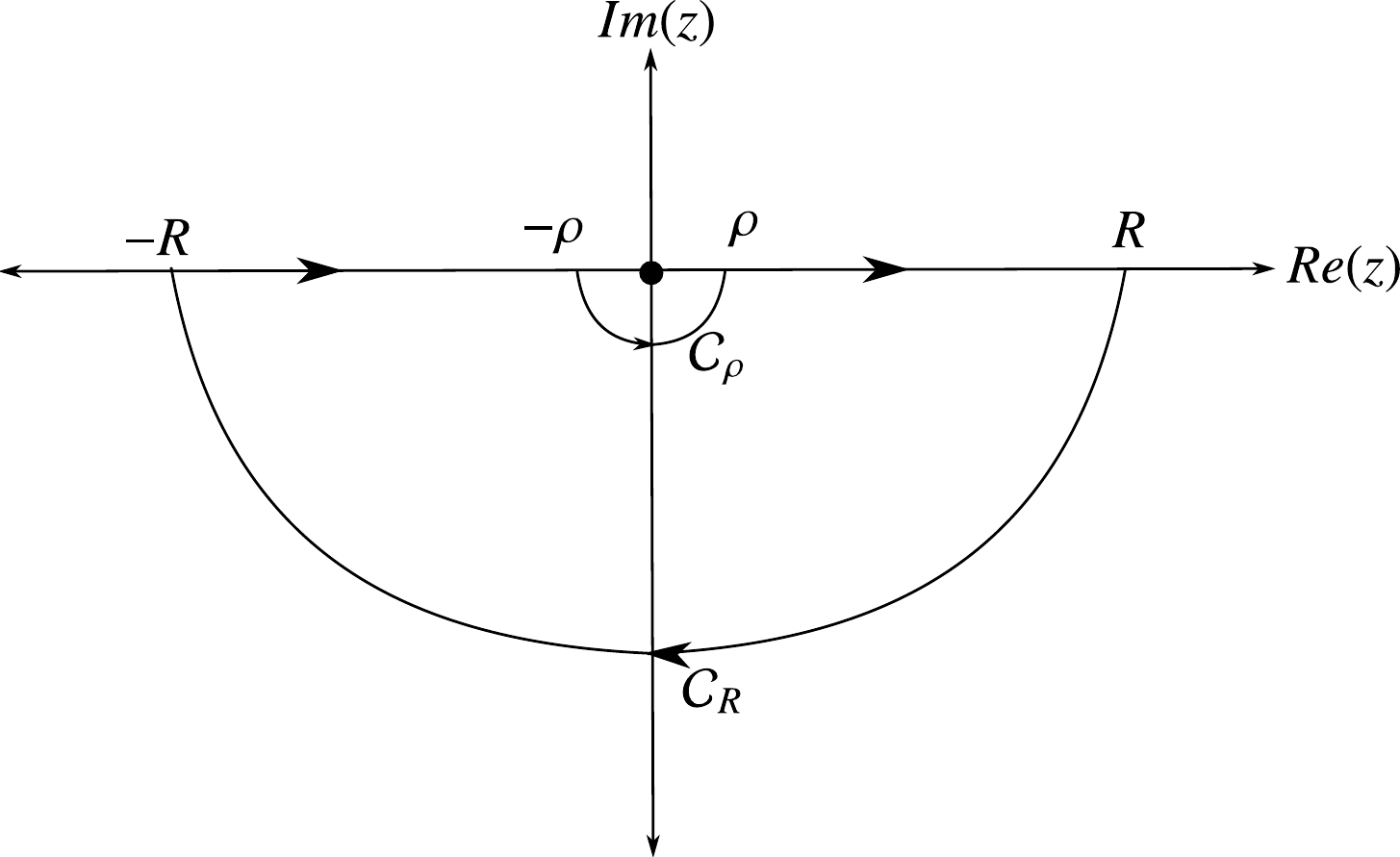}
  \caption{The contour for $\kappa_1 \ge 0 $, and $\kappa_2 \ge 0 $ is shown. In the opposite case, the contour can be chosen in the upper-half plane.}
  \label{contour}
\end{figure}

We will evaluate the following integral
\be
I_n \equiv \int_{-\infty}^{\infty}  d x \,  f_n(x) \equiv \int_{-\infty}^{\infty}  d x \,  \frac{\exp\left[  -i \left( \kappa_1 x  - \dfrac{\kappa_2}{x} \right) \right]}{x^n} \, ,
\ee
with  $\kappa_1 \ge 0 $, and $\kappa_2 \ge 0 $. Accordingly, we consider the corresponding complex integral along the contour shown in Fig.\ref{contour}
\be
\oint_\mathcal{C}  d z \,  f_n(z) = \oint_\mathcal{C}  d z \,  \frac{\exp\left[  -i \left( \kappa_1 z  - \dfrac{\kappa_2}{z} \right) \right]}{z^n} = 0 \, ,
\ee
where the second equality is due to that there is no pole inside the contour. The complex integral can be decomposed as
\bal
& \oint_\mathcal{C}  d z \,  f_n(z)  = I_n + \lim_{\rho \to 0^{+}} \int_{\mathcal{C}_\rho} dz \,  f_n(z)+ \lim_{R \to \infty} \int_{\mathcal{C}_R} dz \,  f_n(z) = 0 \, ,
\eal
then the integral $I_n$ can be given by
\bal
\nonumber I_n  & = -  \lim_{\rho \to 0^{+}} \left[ \int_0^\pi \frac{i d \theta}{\rho^{n-1} e^{i (n-1)(\theta + \pi)}} \exp\left( i \kappa_1 \rho \cos(\theta) - \kappa_1 \rho \sin(\theta) - \frac{i \kappa_2}{\rho} \cos(\theta) -\frac{\kappa_2}{\rho} \sin(\theta) \right) \right] \\
& +  \lim_{R \to \infty} \left[ \int_0^\pi \frac{i d \theta}{R^{n-1} e^{-i (n-1)\theta}} \exp\left( -i \kappa_1 R \cos(\theta) - \kappa_1 R \sin(\theta) + \frac{i \kappa_2}{R} \cos(\theta) -\frac{\kappa_2}{R} \sin(\theta) \right) \right] \, .
\eal
We first observe that when $n=1$, the integral vanishes, $I_1 =0 $. In the case of $n=0$, the integral along the contour $\mathcal{C}_\rho$ vanishes, and the integral $I_0$ becomes
\be
I_0 = \lim_{R \to \infty} \left[ \int_0^\pi i d \theta \,  R e^{- i \theta} \exp\left( -i \kappa_1 R \cos(\theta) - \kappa_1 R \sin(\theta)  \right) \right] \, ,
\ee
which vanishes as long as $\kappa_1 \ne 0$. In the case of $\kappa_1 = 0$, the integral yields $I_0 = \lim_{R \to \infty} 2 R$, which goes to infinity, as a result $I_0 = \delta(\kappa_1)$. In a similar way, when $n=2$, the integral along the contour $\mathcal{C}_R$ goes to zero, and we have
\be
I_2 =  \lim_{\rho \to 0^{+}} \left[ \int_0^\pi \frac{i d \theta}{\rho} e^{-i \theta} \exp\left( -  \frac{i \kappa_2}{\rho} \cos(\theta) -\frac{\kappa_2}{\rho} \sin(\theta) \right) \right] \, ,
\ee
which is zero for $\kappa_2 \ne 0$, otherwise $I_2 = \lim_{\rho \to 0^{+}} 2/\rho $, i.e., $I_2 = \delta(\kappa_2)$. To sum up, we conclude
\be
 \int_{-\infty}^{\infty}  d x \,  \frac{\exp\left[  -i \left( \kappa_1 x  - \dfrac{\kappa_2}{x} \right) \right]}{x^n} = \begin{cases} 
      \delta(\kappa_1) & n = 0 \\
      0 & n = 1 \\
      \delta(\kappa_2) & n = 2
   \end{cases} \, .
\ee

We note that in the case of $\kappa_1 \le 0 $, and $\kappa_2 \le 0 $, which also satisfies $\kappa_1 \kappa_2 \ge 0$, one can chose the integration contour in the upper-half plane of the complex plane, see for a detailed discussion Ref.~\cite{boca2010completeness}.

\bibliography{yakaboylu_bibliography}

\end{document}